\documentclass[a4paper]{jpconf}
\usepackage[square,sort&compress,numbers]{natbib} 
\usepackage{graphicx,iopams}
\begin{document}
\title{Computer Assisted `Proof' of the Global Existence of Periodic
  Orbits in  the R\"{o}ssler System}  \author{Andr\'{e} E Botha$^1$
  and Wynand Dednam$^{1,2}$}  \address{$^1$Department of Physics,
  University of South Africa, P.O. Box 392, Pretoria 0003, South
  Africa}   \address{$^2$Departamento de Fisica Aplicada Universidad
  de Alicante, San  Vicente del Raspeig, E-03690
  \phantom{$^2$}Alicante, Spain}  \ead{bothaae@unisa.ac.za}
\begin{abstract}
The numerical optimized shooting method for finding periodic orbits in
nonlinear dynamical systems was employed to determine the existence of
periodic orbits in the well-known R\"{o}ssler system. By optimizing
the period $T$ and the three system parameters, $a$, $b$  and $c$,
simultaneously,  it was found that, for any initial condition
$(x_0,y_0,z_0) \in \Re^3$, there exists at least one set of optimized
parameters corresponding to a periodic orbit passing through $
(x_0,y_0,z_0)$. After a discussion of this result it was concluded
that its analytical proof may present an interesting new mathematical
challenge.
\end{abstract}
\section{Introduction}
The three-dimensional R\"{o}ssler system (RS) was originally conceived
as a simple  prototype for studying chaos \cite{ros76}. With only one
nonlinear term it can be  thought of as a simplification of the
well-known  Lorenz system \cite{lor63} and as a minimal model for
continuous-time chaos  \cite{gas05}. 

Over the past 38 years the RS has been studied extensively, both in
its own right
\cite{gar85,ter99,lli02,zha04,gal06,cas07,alg07,lli07,bar09,bar11,lim11,tud12}
and as a model, to illustrate various nonlinear phenomena and
different types of chaos \cite{pil99,wil09,gal10,loz13}.
Surprisingly, despite the literally thousands of scientific articles
that have been devoted to the RS, it still poses several open
questions. For example, although the conjecture by Leonov
\cite{leo12},  about  the Lyapunov dimension of the RS (and two other
types of R\"{o}ssler  systems), has recently been verified numerically
\cite{kuz14}, the exact Lyapunov dimension for the RS is not known
analytically.  Leonov \cite{leo14} has recently also provided an
analytical  estimate of the attractor dimension in the RS, however,
similar estimates for slightly more general R\"{o}ssler systems remain
a difficult open problem. There is also an ongoing discussion about
whether or not the RS  can indeed  produce a strange attractor, in the
same sense as that of the Lorenz system \cite{zgl97,loz13}. 

More pertinent to the present study are the recent articles on the
existence or non-existence and/or classification of the periodic
orbits of the RS, in terms of its three control parameters $a$, $b$
and and $c$. Starkov and Starkov \cite{sta07} have shown that, when
the parameters satisfy the condition $c^{2}-4ab<0$, no periodic
solutions can exist. They have also proved that, when $ac<0$ ($ac>0$)
and $ab>0$, the periodic orbits exist entirely  in the negative
(positive) half space, i.e. $z<0$ ($z>0$).

There have been several attempts to demonstrate the existence of
orbits more generally; such as those by Genesio and Ghilardi
\cite{gen05}, in which the existence of the quasi-periodic orbits in
third-order systems like those of R\"{o}ssler, Lorenz and Chua {\em et
  al}.  \cite{chu86} are proved. More recently, instead of the focus
of such studies being on the unstable orbits, the experimentally
observable chaotic and stable periodic orbits of simple chaotic flows,
including the RS, have been classified very generally in terms of
their control parameters \cite{gal10}.  In the parameter space,
families of stable periodic solutions have been found to self-organize
about the so-called isolated periodicity hubs of the RS. (See
Ref.~\cite{gal10} and the references therein.) 

Another active area of related research is on the integrability of the
RS \cite{hit97,lli02,cha09,lli07,lim11,tud12}. In Ref.~\cite{lli02} it
was shown that Darboux integrability of the RS for various parameter
values leads to surfaces in state space containing periodic
orbits. The bifurcations and routes to chaos in the RS have also been
investigated \cite{gar85,nik04} and there is at least one exhaustive
study on the complete parametric evolution of the system
\cite{bar09}. (Also see Ref.~\cite{bar11} and the references therein.)

The present work developed as a result of using the RS as a test case
for our recently-developed numerical technique (called the optimized
shooting method) for finding periodic solutions in nonlinear dynamical
systems \cite{ded14}. During the course of testing our method we
discovered that it could optimize the system parameters to find at
least one periodic orbit for {\em any} initial condition. At first
this finding was very surprising to us, and we assumed that it was
most likely due to a coding error. We therefore tested the codes more
thoroughly and also experimented with a variety of different
integration schemes and platforms, in order to verify the results
independently. After confirming that our numerical results were indeed
accurate, we turned to the literature to establish if a similar result
had been reported elsewhere for the RS. However, in all of  the
related literature we found, there appeared to be no explicit proof
(or mention) of the result we had obtained. Hence we consider it
worthy of a separate brief report.

We are certainly not the first to use numerical computations to
`prove', i.e.  motivate, the existence of certain analytical
properties.  Pilarczyk \cite{pil99}, for example, provides a computer
assisted `proof' of the existence of a periodic orbit in the
RS. Wilczak and Zgliczy\'{n}ski \cite{wil09} have also made use of a
numerical method to `prove' the existence of two period-doubling
bifurcations in R\"{o}ssler's system, as well as the existence of a
branch of  period two points connecting them.

The remaining material in this article is organized as follows. In
Sec. 2 we provide a brief description of the optimized shooting
method. In Sec. 3 we state the main result in the form of a conjecture
and  describe the procedure that was followed in order to motivate
it. Section 4 ends  with a short discussion and conclusion.
\section{Optimized shooting method} 
Today there are a variety of numerical methods available for finding
the periodic orbits of nonlinear dynamical systems \cite{aba11}. For
the present purpose we make use of the  optimized shooting method
\cite{ded14}, which originally enabled us to discover the reported new
property of the periodic orbits in the RS \cite{ros76}.   Briefly,
this method makes use of Levenberg-Marquardt optimization to find the
periodic orbits by minimizing a residual (or error function)
\cite{lev44,mar63}. 

To apply the method to the RS, we re-write the system equations as    
\begin{equation}
\dot{x} = T\left(-y - z\right), \; \; \;  \dot{y} = T\left( x +
ay\right), \; \; \; \dot{z} = T\left(b + z\left( x-c\right)
\right),  \label{eq1}
\end{equation}
where $T$ is the (as yet) unknown period of the desired solution and
the overdot indicates differentiation with respect to the
dimensionless time  $\tau = t/T$. Since one period corresponds to
integration over $\tau$ from zero to one,  we define the residual as 
\begin{equation}
\mathbf{R} = \left( \mathbf{x} \left( 1\right) -\mathbf{x}\left(
0\right) ,\mathbf{x}\left( 1 + \Delta \tau \right)
-\mathbf{x}\left(\Delta \tau \right)),\ldots, \mathbf{x}\left(
1+p\Delta \tau \right) -\mathbf{x} \left( p \Delta \tau \right)
\right) ,  \label{eq2}
\end{equation}
where $\Delta \tau$ is a fixed integration step size, $p =
0,1,2,\ldots$  and $\mathbf{x} \equiv \left( x,y,z \right)$. The
residual is a function of  the initial conditions, period, and three
system parameters; since it depends on these through solution of
Eq.~(\ref{eq1}). Furthermore, since $\mathbf{x}(\tau) =
\mathbf{x}(\tau+1)$ for periodic  solutions, it can immediately be
seen that such solutions correspond to  a vanishing residual, i.e. the
periodic solutions are found by optimizing $\mathbf{R}$ to be as close
as possible to zero.

\section{Results}
We state the main result of this work in the form of a conjecture.
\begin{quote}
{\bf Conjecture}: For any initial condition $(x_0,y_0,z_0) \in \Re^3$,
there exists real parameters, $a$, $b$ and $c$, for Eq. (\ref{eq1}),
such that its solution is periodic.
\end{quote}
For initial conditions on plane $z=0$,  inspection of Eq. (\ref{eq1})
shows that the conjecture is trivially satisfied: for $a=b=c=0$ the
solutions are circles, with periods equal to $2\pi$. The analytical
proof of the integrability of the system, for this case, is given in
Refs. \cite{zha04} and \cite{lli07}.

Off the $z=0$ plane the validity of the conjecture is established
numerically by performing optimization of the four parameters, $T$,
$a$, $b$ and $c$, for sets of randomly selected initial conditions. To
facilitate the selection of these initial conditions it is convenient
to  rewrite Eq. (\ref{eq1}) in  terms of scaled (primed) coordinates,
defined by $x = \alpha x^{\prime} $,  $y = \alpha y^{\prime} $, and $z
= \alpha z^{\prime} $. Here $\alpha \ge 1$ is the scale factor. In
terms of the primed coordinates Eq. (\ref{eq1}) is given by 
\begin{equation}
\dot{x}^{\prime} =  T\left(-y^{\prime} - z^{\prime}\right), \; \; \;
\dot{y}^{\prime} =  T\left(x^{\prime}+ay^{\prime}\right), \; \; \;
\dot{z}^{\prime} =  T\left(b/\alpha + \alpha z^{\prime} \left(
x^{\prime} - c/\alpha \right) \right) . \label{eq3}
\end{equation}
For a given $\alpha$, the procedure for selecting the initial
conditions then consisted of generating $100$ distinct random points,
within or on the unit sphere. For each initial condition  the
optimized shooting method  was used to find the  four parameters that
produce a periodic orbit.  As an initial guess for the parameters, we
chose $(T,a,b,c) = (5,1,1,1)$. Starting  from $\alpha = 1$, the value
of  the scale factor was increased, in steps  of $1$, up to the
maximum value of $100$. This procedure produced a total of
$100\times100$ random initial conditions. All $10^4$ initial
conditions were successfully optimized, i.e. in each case the
magnitude of $\mathbf{R}$ was successfully optimized to be below the
set numerical  tolerance ($\left| \mathbf{R} \right| < 10^{-12}$ in
this calculation).    

Table \ref{tab1} lists the optimized parameters for $15$ arbitrarily
chosen  periodic orbits. 
\begin{table}[h!]
\caption{Fifteen of the $10^4$ periodic orbits found for the
  R\"{o}ssler system via the optimized  shooting method. The column
  headings are as follows: $n$ is the orbit number (used to refer to
  the orbit in the main text), $\alpha$ is the scale parameter
  introduced in the main text (see Eq. \ref{eq3}), $x^{\prime}_0$,
  $y^{\prime}_0$ and $z^{\prime }_0$ are the primed coordinates of the
  randomly selected initial conditions, $T$ is the period, and the
  last three columns give the optimized system
  parameters. \label{tab1}}
\begin{center}
\begin{tabular}{ccccccccc}
\br $n$ & $\alpha $ & $x_{0}^{\prime }$ & $y_{0}^{\prime }$ &
$z_{0}^{\prime }$ & $T$ & $a$ & $b$ & $c$ \\ \mr \multicolumn{1}{l}{1}
& \multicolumn{1}{l}{1} & \multicolumn{1}{r}{-0.322} &
\multicolumn{1}{r}{0.283} & \multicolumn{1}{r}{-0.827} &
\multicolumn{1}{r}{ 6.285} & \multicolumn{1}{r}{-0.01832876699} &
\multicolumn{1}{r}{ -37.861974659} & \multicolumn{1}{r}{45.448602039}
\\  \multicolumn{1}{l}{2} & \multicolumn{1}{l}{1} &
\multicolumn{1}{r}{0.083} &  \multicolumn{1}{r}{0.498} &
\multicolumn{1}{r}{0.756} & \multicolumn{1}{r}{ 6.282} &
\multicolumn{1}{r}{0.01659280123} & \multicolumn{1}{r}{34.3119795334 }
& \multicolumn{1}{r}{45.496814976} \\  \multicolumn{1}{l}{3} &
\multicolumn{1}{l}{1} & \multicolumn{1}{r}{0.271} &
\multicolumn{1}{r}{-0.033} & \multicolumn{1}{r}{-0.492} &
\multicolumn{1}{r}{ 6.284} & \multicolumn{1}{r}{-0.01074867143} &
\multicolumn{1}{r}{ -22.256338664} & \multicolumn{1}{r}{45.495734090}
\\  \multicolumn{1}{l}{4} & \multicolumn{1}{l}{1} &
\multicolumn{1}{r}{-0.843} &  \multicolumn{1}{r}{-0.154} &
\multicolumn{1}{r}{-0.338} & \multicolumn{1}{r}{ 6.283} &
\multicolumn{1}{r}{-0.00756927443} & \multicolumn{1}{r}{
  -15.661740755} & \multicolumn{1}{r}{45.483293583}
\\  \multicolumn{1}{l}{5} & \multicolumn{1}{l}{1} &
\multicolumn{1}{r}{0.920} &  \multicolumn{1}{r}{-0.101} &
\multicolumn{1}{r}{-0.062} & \multicolumn{1}{r}{ 6.285} &
\multicolumn{1}{r}{-0.00133598606} & \multicolumn{1}{r}{
  -2.7619826425} & \multicolumn{1}{r}{45.463987246}
\\  \multicolumn{1}{l}{6} & \multicolumn{1}{l}{10} &
\multicolumn{1}{r}{0.274} &  \multicolumn{1}{r}{-0.843} &
\multicolumn{1}{r}{-0.019} & \multicolumn{1}{r}{ 6.280} &
\multicolumn{1}{r}{-0.00346652925} & \multicolumn{1}{r}{
  -9.6387456311} & \multicolumn{1}{r}{53.299901841}
\\  \multicolumn{1}{l}{7} & \multicolumn{1}{l}{10} &
\multicolumn{1}{r}{-0.431} & \multicolumn{1}{r}{0.429} &
\multicolumn{1}{r}{-0.132} &  \multicolumn{1}{r}{6.287} &
\multicolumn{1}{r}{-0.02714795455} & \multicolumn{1}{r}{-75.519091787}
& \multicolumn{1}{r}{52.954713982} \\  \multicolumn{1}{l}{8} &
\multicolumn{1}{l}{10} & \multicolumn{1}{r}{-0.762} &
\multicolumn{1}{r}{0.094} & \multicolumn{1}{r}{-0.023} &
\multicolumn{1}{r}{6.283} & \multicolumn{1}{r}{-0.00504377136} &
\multicolumn{1}{r}{-13.929997631} & \multicolumn{1}{r}{52.959008822}
\\  \multicolumn{1}{l}{9} & \multicolumn{1}{l}{10} &
\multicolumn{1}{r}{0.721} &  \multicolumn{1}{r}{0.212} &
\multicolumn{1}{r}{0.532} & \multicolumn{1}{r}{ 6.303} &
\multicolumn{1}{r}{0.09058831072} & \multicolumn{1}{r}{242.356061832 }
& \multicolumn{1}{r}{52.926271854} \\  \multicolumn{1}{l}{10} &
\multicolumn{1}{l}{10} & \multicolumn{1}{r}{0.212} &
\multicolumn{1}{r}{-0.008} & \multicolumn{1}{r}{-0.946} &
\multicolumn{1}{r}{6.412} & \multicolumn{1}{r}{-0.18855060455} &
\multicolumn{1}{r}{-483.78516338} & \multicolumn{1}{r}{53.074125658}
\\  \multicolumn{1}{l}{11} & \multicolumn{1}{l}{100} &
\multicolumn{1}{r}{0.370} & \multicolumn{1}{r}{0.525} &
\multicolumn{1}{r}{-0.489} &  \multicolumn{1}{r}{9.221} &
\multicolumn{1}{r}{-0.72151348243} & \multicolumn{1}{r}{-2913.5197570}
& \multicolumn{1}{r}{96.641076651} \\  \multicolumn{1}{l}{12} &
\multicolumn{1}{l}{100} & \multicolumn{1}{r}{0.465} &
\multicolumn{1}{r}{-0.177} & \multicolumn{1}{r}{0.111} &
\multicolumn{1}{r}{6.506} & \multicolumn{1}{r}{0.27456932042} &
\multicolumn{1}{r}{1.25094211568} & \multicolumn{1}{r}{35.351207642}
\\  \multicolumn{1}{l}{13} & \multicolumn{1}{l}{100} &
\multicolumn{1}{r}{0.045} & \multicolumn{1}{r}{-0.881} &
\multicolumn{1}{r}{0.206} & \multicolumn{1}{r}{1.670} &
\multicolumn{1}{r}{0.28422944418} & \multicolumn{1}{r}{0.06378881294}
& \multicolumn{1}{r}{25.148204118} \\  \multicolumn{1}{l}{14} &
\multicolumn{1}{l}{100} & \multicolumn{1}{r}{0.448} &
\multicolumn{1}{r}{-0.071} & \multicolumn{1}{r}{0.533} &
\multicolumn{1}{r}{7.016} & \multicolumn{1}{r}{0.41997047657} &
\multicolumn{1}{r}{5480.91338691} & \multicolumn{1}{r}{148.08093025}
\\  \multicolumn{1}{l}{15} & \multicolumn{1}{l}{100} &
\multicolumn{1}{r}{0.177} & \multicolumn{1}{r}{-0.786} &
\multicolumn{1}{r}{0.321} & \multicolumn{1}{r}{3.795} &
\multicolumn{1}{r}{0.35076508274} & \multicolumn{1}{r}{0.24268923009}
& \multicolumn{1}{r}{27.447807055} \\ \br
\end{tabular}
\end{center}
\end{table}
For ease of presentation, the randomly chosen initial conditions were
first rounded off to three decimal places, before applying the
optimized shooting method. Thus the initial conditions, as listed in
columns 3-5 of Table \ref{tab1}, are exact. On the other hand the
periods listed in column 6 are given approximately to three decimal
places. Since they are not required to reconstruct the orbits, their
full  11 digit accuracy is not required here.

Two other features of Table \ref{tab1} are also worth
mentioning. First, we notice that the signs of the parameters $a$ and
$b$ are always the same as that of $z_{0}^{\prime }$. This is a
property of all the found orbits and it confirms the analytical result
proved by  Starkov and Starkov \cite{sta07}: when $ac<0$ ($ac>0$) and
$ab>0$, the periodic orbits exist entirely  in the negative (positive)
half space.  Second, in Ref. \cite{sta07} it was also shown that for
$c^{2}-4ab<0$, no periodic orbits can exist. In Table 1, and indeed
for all the found orbits,  $c^{2}-4ab$ is positive. Thus our results
are consistent with the known classifications for periodic orbits in
the RS.

\begin{figure}[h]
\centering   \includegraphics[width=0.46\textwidth]{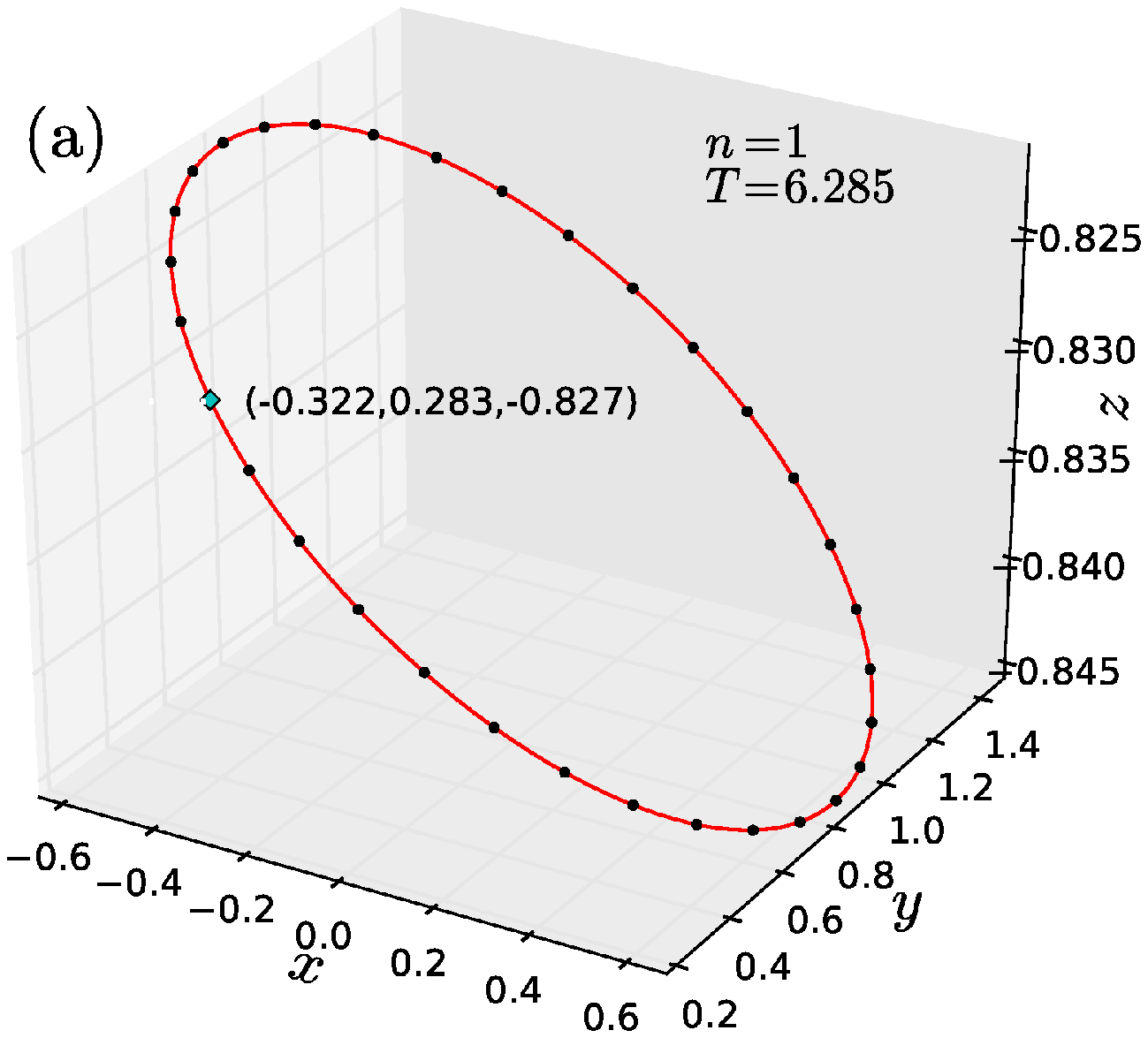}
\includegraphics[width=0.46\textwidth]{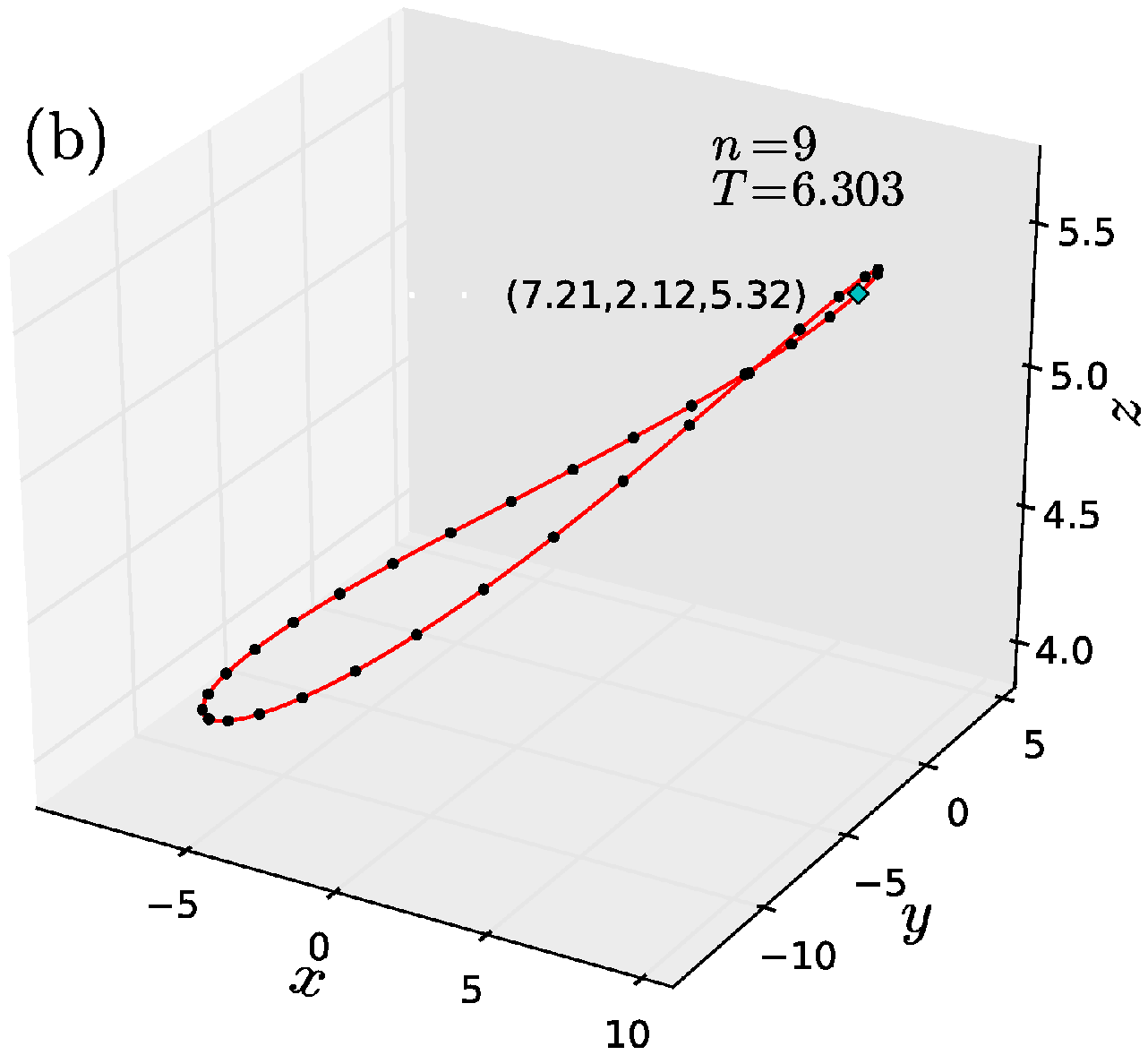}

\includegraphics[width=0.46\textwidth]{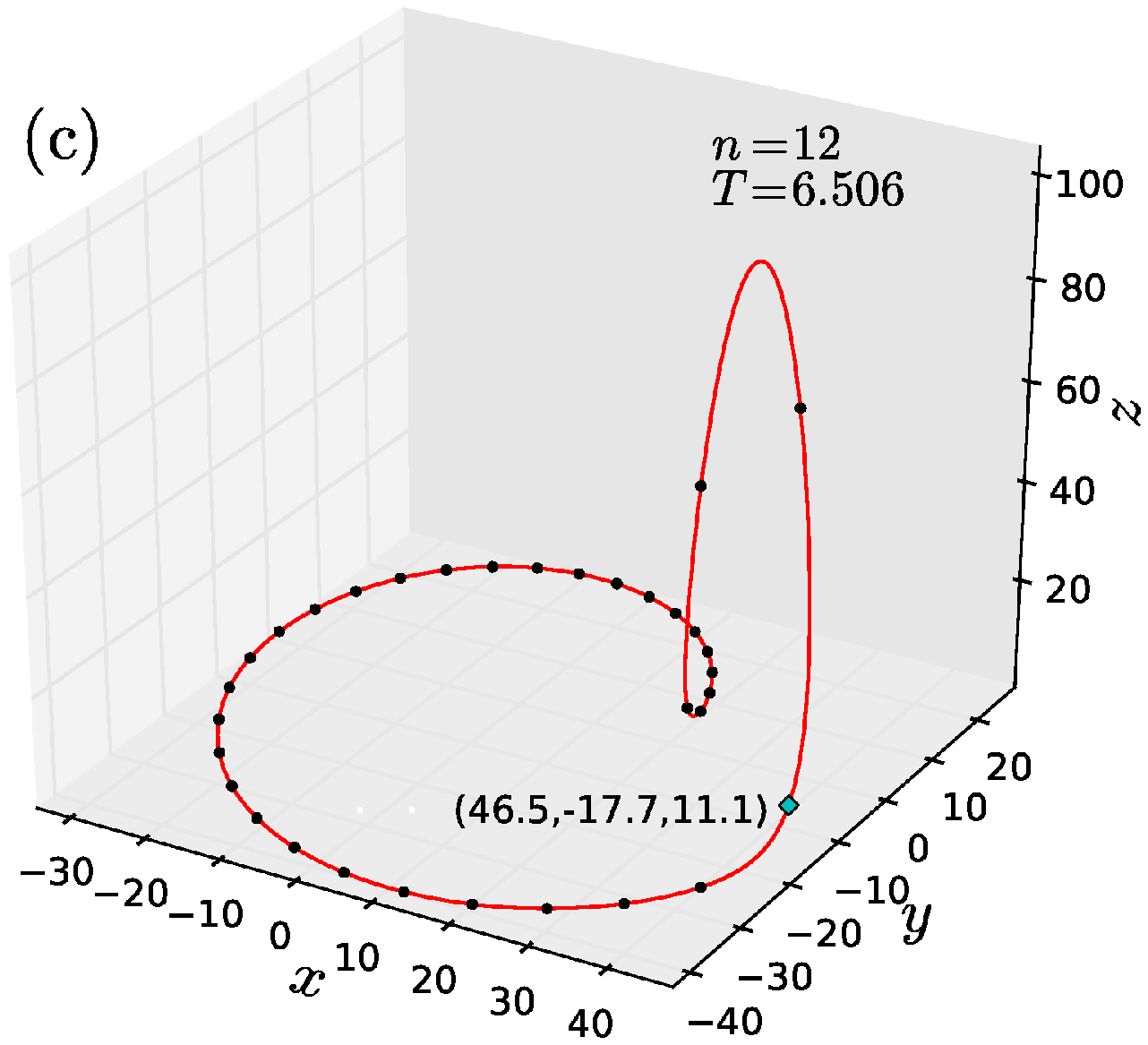}
\includegraphics[width=0.46\textwidth]{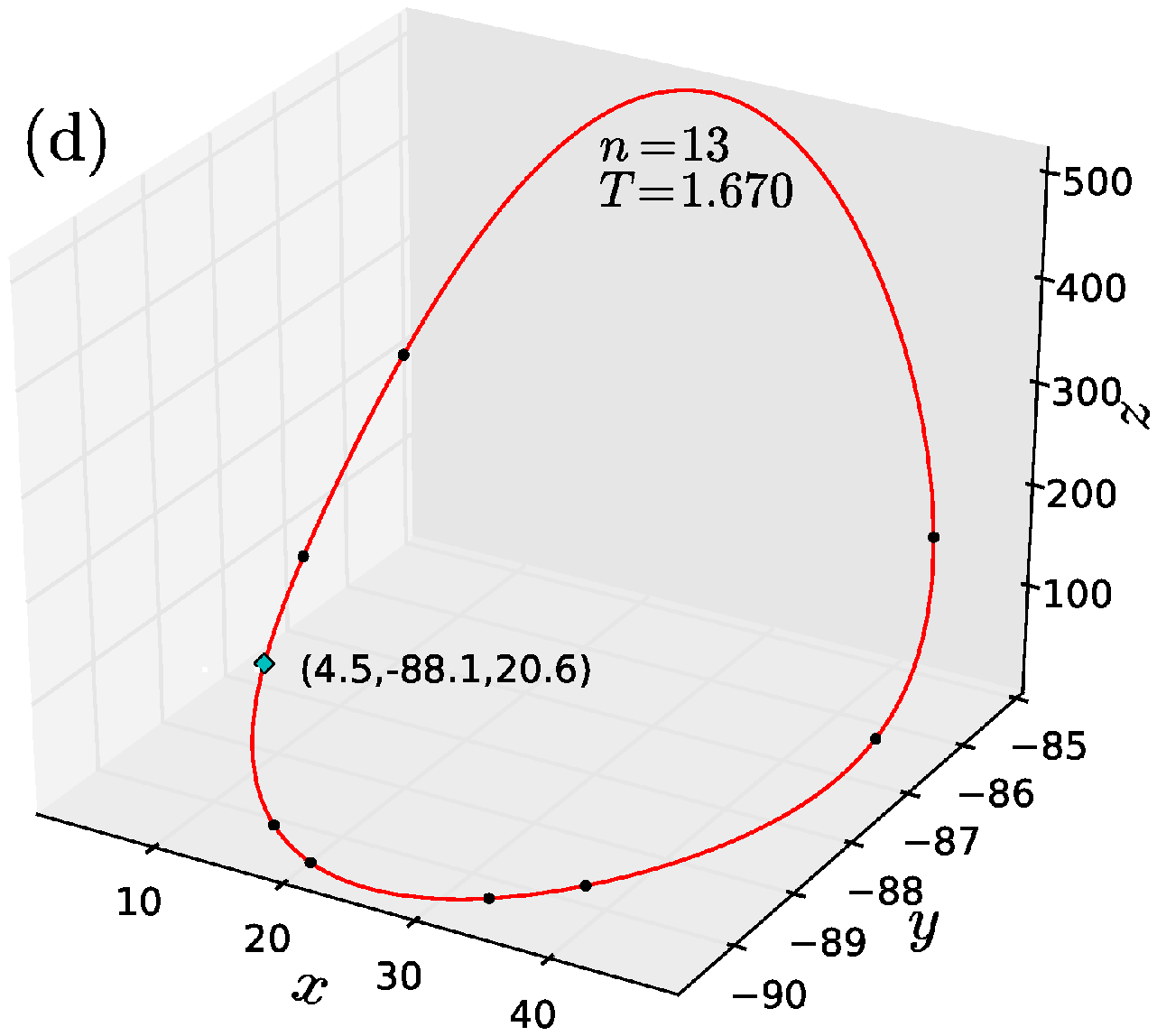}

\caption{Phase portraits of four arbitrarily selected periodic orbits,
  chosen from  Table  \ref{tab1}. The initial condition for each orbit
  is indicated by a blue square marker. Black filled circular markers
  indicate points along the orbit, every 200 integration time steps.
  Since a fixed time step of $\Delta \tau = 1/1024$ was used, the
  circular markers provide an indication of the velocity
  $(\dot{x}(\tau),\dot{y}(\tau),\dot{z}(\tau))$ along each
  orbit. \label{fig1}}
\end{figure}

In total more than 100 spot checks were made on randomly selected
orbits taken from the set of $10^4$. Such checks were made by plotting
and visually inspecting the selected orbits. Figure~\ref{fig1} shows
the plots for four orbits that are listed in Table \ref{tab1}, i.e.
orbits numbers $1, 9, 12$ and $13$.  In Fig.~\ref{fig1}(a) it can be
seen that orbit one is confined to lie within the negative half space
$\Re^3-\{z \ge 0\}$, while orbits 9, 12 and 13, shown in
Figs.~(b)-(d), occur in the positive half space $\Re^3-\{z \le 0
\}$. This observation is consistent with the form of
Eq. (\ref{eq1}). If  periodic orbits existed that crossed from one
half space into the other, they would have  to intersect the plane
$z=0$ more than once. Moreover, at $z=0$  the sign of $\dot{z}$ would
have to be positive (negative) in going from the negative (positive)
half space to the positive (negative) half space. However, the form of
Eq. (\ref{eq1}) clearly excludes this possibility, since for  $z=0$
the sign of $\dot{z}$ is determined by the parameter $b$. Therefore,
for a given set of parameters, the sign of $\dot{z}$ at $z=0$ cannot
change. Hence, the non-trivial periodic orbits are confined to lie
entirely within one half space or the other.   

Finally, as an alternative way (i.e. without using the residual) of
estimating  the accuracy to which the  optimized shooting method is
capable of determining the periodic solutions,  several initial
conditions with $z=0$ were also examined. In all such cases the
magnitudes of the largest optimized system parameters were found to be
less than $10^{-11}$, with the expected period of $2\pi$ also
given accurately to eleven decimal places. 

\section{Discussion and Conclusion}
We have established a computer assisted `proof' of the global
existence of periodic solutions in the R\"{o}ssler system. By global
we mean that, for any initial condition $(x_0,y_0,z_0) \in \Re^3$, our
numerical method was able to optimize simultaneously the period $T$
and system parameters, $a$, $b$  and $c$, in order to find at least
one periodic solution passing through the initial condition. We have
stated this result as a conjecture, as opposed to a proposition,
because the latter term is generally reserved for mathematical results
that can  be proved rigorously. The term `proof', which we have 
written consistently in single quotes, should thus be understood to mean a 
numerical justification/motivation, rather than a mathematical proof. 
This is also the sense in which the same term is used elsewhere in the 
literature on computer assisted proofs. (See, for example, Refs. 
\cite{pil99} and \cite{wil09}.)

It is interesting to note that, in some cases, more than one solution
passing though a given initial condition could be found, i.e. two or
more distinct sets of the optimized parameters could be found for a
given initial condition. For example, in the case of orbit number 12
(see Fig.~\ref{fig1}(c) and Table~\ref{tab1}), there also exists a
different  period orbit passing through the same initial point, with
precisely twice the period, i.e. $T = 13.0116560813$, at slightly
different system parameters. This multiplicity, of period doubled
orbits all passing through the same point, offers an interesting new
possibility. Whereas conventionally the universal period doubling
route to chaos (the so-called Feigenbaum scenario \cite{fei78,fei79})
is usually achieved by varying one system parameter at a time, we see
here that it may also be possible (by following a very specific (three
dimensional) path in the parameter space) to obtain a special sequence
of period doubling orbits {\em which all pass through the same point
  in the phase space}. It remains to be seen whether such a special
sequence would also lead to chaos and follow Feigenbaum's universal
scaling laws.
 
In conclusion, we have formulated a conjecture about the global
existence of periodic solutions in R\"{o}ssler's system. Our
conjecture is numerically supported (i.e. `proved') and it postulates
the existence of periodic orbits passing through {\em any} point in
the  phase space, for a suitable choice of the system
parameters. While it is in principle possible to establish this result
analytically, to date there  does not appear to be any mention in  the
literature of either the result itself or its proof.  Unfortunately we
were not able to construct  such an analytical proof
ourselves. However, in view of the compelling computational evidence
which we have provided, we hope that some theoretically inclined
readers may find it interesting to devise a rigorous mathematical
proof of our conjecture. 

\ack The financial assistance of the National Research Foundation
of South Africa towards this research is hereby acknowledged by WD.

\bibliographystyle{iopart-num}
\bibliography{chaos}

\providecommand{\newblock}{}
\begin{thebibliography}{10}
\expandafter\ifx\csname url\endcsname\relax
  \def\url#1{{\tt #1}}\fi
\expandafter\ifx\csname urlprefix\endcsname\relax\def\urlprefix{URL }\fi
\providecommand{\eprint}[2][]{\url{#2}}

\bibitem{ros76}
R{\"o}ssler O~E 1976 {\em Phys. Lett. A\/} {\bf 57} 397

\bibitem{lor63}
Lorenz E~N 1963 {\em J. Atmos. Sci.\/} {\bf 20} 130

\bibitem{gas05}
Gaspard P 2005 {\em Encyc. Nonlin. Sci.\/} ed Scott A (New York: Routledge) p
  808

\bibitem{gar85}
Gardini L 1985 {\em Nuovo Cimento\/} {\bf 89B} 139

\bibitem{ter99}
Teryokhin M~T and Panfilova T~L 1999 {\em Russian Mathematics\/} {\bf 43} 66

\bibitem{lli02}
Lliubre J and Zhang X 2002 {\em Int. J. Bifurcation Chaos\/} {\bf 12} 421

\bibitem{zha04}
Zhang X 2004 {\em Int. J. Bifurcation Chaos\/} {\bf 14} 4275

\bibitem{gal06}
Galias Z 2006 {\em Int. J. Bifurcation Chaos\/} {\bf 16} 2873

\bibitem{cas07}
Castro V {\em et~al.\/} 2007 {\em Int. J. Bifurcation Chaos\/} {\bf 17} 965

\bibitem{alg07}
Algaba A {\em et~al.\/} 2007 {\em Int. J. Bifurcation Chaos\/} {\bf 17} 1997

\bibitem{lli07}
Lliubre J and Valls C 2007 {\em Int. J. Bifurcation Chaos\/} {\bf 17} 3289

\bibitem{bar09}
Barrio R, Blesa F and Serrano S 2009 {\em Physica D\/} {\bf 238} 1087

\bibitem{bar11}
Barrio R, Blesa F, Dena A and Serrano S 2011 {\em Computers and Mathematics
  with Applications\/} {\bf 62} 4140

\bibitem{lim11}
Lima M~F~S and Llibre J 2011 {\em J. Phys. A: Math. Theor.\/} {\bf 44} 365201

\bibitem{tud12}
Tudoran R~M and Girban A 2012 {\em J. Math. Phys.\/} {\bf 53} 052701

\bibitem{pil99}
Pilarczyk P 1999 {\em Topological Methods in Nonlinear Analysis\/} {\bf 13} 365

\bibitem{wil09}
Wilczak D and Zgliczy\'{n}ski P 2009 {\em Found. Comput. Math.\/} {\bf 9} 611

\bibitem{gal10}
Gallas J~A~C 2010 {\em Int. J. Bifurcation Chaos\/} {\bf 20} 197

\bibitem{loz13}
Lozi R 2013 {\em Topology and Dynamics of Chaos\/} ed Letellier C and Gilmore R
  (Singapore: World Scientific)

\bibitem{leo12}
Leonov G 2012 {\em J. Appl. Math. Mech.\/} {\bf 76} 129

\bibitem{kuz14}
Kuznetsov N, Mokaev T~N and Vasilyev P~A 2014 {\em Comm. in Nonl. Sci and Num.
  Simul.\/} {\bf 19} 1027--1034

\bibitem{leo14}
Leonov G~A 2014 {\em Doklady Mathematics\/} {\bf 89} 1--3

\bibitem{zgl97}
Zgliczy\'{n}ski P 1997 {\em Nonlinearity\/} {\bf 10} 243

\bibitem{sta07}
Starkov K~E and Starkov K~K 2007 {\em Chaos, Solitons and Fractals\/} {\bf 33}
  1445

\bibitem{gen05}
Genesio R and Ghilardi C 2005 {\em Int. J. Bifurcation Chaos\/} {\bf 15} 3165

\bibitem{chu86}
Chua L, Komuro M and Matsumoto T 1986 {\em IEEE Trans. Circuits Syst.\/} {\bf
  33} 1072

\bibitem{cha09}
Chandrasekar V, Senthilvelan M and Lakshmanan M 2009 {\em Proc. R. Soc. A\/}
  {\bf 465} 585

\bibitem{hit97}
Hitchin N 1997 {\em Integrable systems: An Introduction\/} (Oxford: Notes from
  the Mathematical Institute)

\bibitem{nik04}
Nikolov S and Petrov V 2004 {\em Int. J. Bifurcation Chaos\/} {\bf 14} 293

\bibitem{ded14}
Dednam W and Botha A~E 2014 {\em arXiv:1405.5347v1\/}

\bibitem{aba11}
Abad A, Barrio R and Dena A 2011 {\em Phys. Rev. E\/} {\bf 84} 016701

\bibitem{lev44}
Levenberg K 1944 {\em Quart. Appl. Math.\/} {\bf 2} 164

\bibitem{mar63}
Marquardt D~W 1963 {\em SIAM J. Appl. Math.\/} {\bf 11} 431

\bibitem{fei78}
Feigenbaum M~J 1978 {\em Journal of Statistical Physics\/} {\bf 19} 25

\bibitem{fei79}
Feigenbaum M~J 1979 {\em Journal of Statistical Physics\/} {\bf 21} 669

\end{thebibliography}
\end{document}